# Blockchain Based Information Security and Privacy Protection: Challenges and Future Directions using Computational Literature Review

Gauri Shankar
LUT University, Finland
gauri.shankar@lut.fi

Md Raihan Uddin
LUT University, Finland
md.uddin@lut.fi

Saddam Mukta
LUT University, Finland
saddam.mukta@lut.fi

Prabhat Kumar
LUT University, Finland
prabhat.kumar@lut.fi

Shareeful Islam
Anglia Ruskin University, United Kingdom
shareeful.islam@aru.ac.uk

A.K.M. Najmul Islam
LUT University, Finland
najmul.islam@lut.fi

## Abstract

*Blockchain technology is an emerging digital innovation that has gained immense popularity in enhancing individual security and privacy within Information Systems (IS). This surge in interest is reflected in the exponential increase in research articles published on blockchain technology, highlighting its growing significance in the digital landscape. However, the rapid proliferation of published research presents significant challenges for manual analysis and synthesis due to the vast volume of information. The complexity and breadth of topics, combined with the inherent limitations of human data processing capabilities, make it difficult to comprehensively analyze and draw meaningful insights from the literature. To this end, we adopted the Computational Literature Review (CLR) to analyze pertinent literature's impact and topic modelling using the Latent Dirichlet Allocation (LDA) technique. We identified 10 topics related to security and privacy and provided a detailed description of each topic. From the critical analysis, we have observed several limitations, and several future directions are provided as an outcome of this review.*

## 1. Introduction

Blockchain, a distributed ledger system, has garnered significant attention from industry experts and academics as a potential solution for developing secure and privacy-oriented information systems (IS). The convergence of various cutting-edge technologies has brought new challenges to protecting information. security and users' privacy (Hughes et al., 2019). In today's world, data sharing and analyzing personal information are common practices in interdomain communication systems. Therefore, any security breach can lead to identity theft, confidentiality issues, financial losses, business damage, and loss of customer trust (Wylde et al., 2022). However, Blockchain integration addresses these issues as its decentralized and tamper-proof ledger provides enhanced security and privacy through cryptographic verification and consensus mechanisms (Puthal et al., 2018).

In recent years, significant advancements have been made in utilizing blockchain technology to enhance the privacy and security of IS. Researchers have developed various methodologies integrating advanced security and privacy protection protocols like Zero-Knowledge Proof (ZKP), Multiparty Computations (MPC), and Fully Homomorphic Encryption to address security and privacy issues on the blockchain (Ma et al., 2023; Waheed et al., 2020; Wang et al., 2023). The research on advanced security and privacy protection protocols has led to an increasing number of articles being published exclusively on IS, highlighting the challenges in developing security and privacy-oriented blockchains. To develop an enhanced version of blockchain architecture, it is essential to analyze the current literature (Homoliak et al., 2020), which helps to identify critical research gaps and formulate research questions that address pertinent issues related to security and privacy on the blockchain.

Furthermore, keeping up-to-date with this field's latest topics and trends is crucial for researchers to advance their work. Scholarly publications on blockchain-based privacy and security systems are significantly increasing. Researchers are doing systematic and bibliometric reviews of previous works to identify challenges, opportunities, and potential solutions. However, biases arise when selecting articles from an extensive collection. To address this challenge, we use the CLR topic modeling method (Antons et al., 2023), a technique used to uncover the hidden thematic structure from a large corpus of literature. CLR enables the conduct of a more comprehensive and unbiased analysis of the literature of around 3,904 articles. This analysis helps us identify the top trending topic of privacy and security using blockchain in IS. The CLR helps us recognize critical future research questions for privacy-preserving blockchain architecture. We found the absence of CLR methods on blockchain during previous work reviews for security and privacy studies. This is the key motivation for our research, as we saw the potential for CLR to enhance the quality and objectivity of such reviews in the field of information security and privacy using blockchain for IS.

This article is structured as follows: First, we discuss CLR in detail about blockchain-related works in IS. Then, we explain the methods and tools used in this work, which include impact and content analysis to identify the topics. The next section provides analysis results with visualizations. Finally, we will discuss each topic in detail and describe it. Based on this discussion, we developed research questions.

## 2. Background

### 2.1 Computational Literature Review

There has been a significant increase in the large amount of literature, which challenges scholars in identifying specific research domains. Scholars use traditional literature review methods such as systematic and bibliometric literature reviews to minimize bias in article selection (Thilakaratne et al., 2019). Systematic literature reviews (SLRs) (Nightingale, 2009) have emerged as the preferred method for rigorous and reproducible literature analysis. Further, researchers find new techniques and tools that combine computational techniques with the robustness of SLR. This technique increases the number of performances and reduces the complexity of the review process. One such method is CLR (Antons et al., 2023), which uses advanced machine learning (ML) algorithms to analyze large amounts of textual data systematically. The automated process of CLR accelerates the review process and increases the scope of insights gained from the literature. CLR is based on topic modelling algorithms, such as Latent Dirichlet Allocation (LDA) (Blei et al., 2003), which identifies latent themes within a corpus of documents. The LDA topic modeling works on the unsupervised probabilistic statistical model to identify the hidden topics in the corpus from the content of the articles (Blei et al., 2003).

To perform CLR, researchers follow a structured framework involving six key steps:
1. Begin with a conceptual goal that is the researcher's motivation to perform the review.
2. Operationalising the CLR helps identify the boundaries for selecting the content and the area of the work.
3. Choose a computational technique regarding the suitability of the corpus of the document.
4. Perform the content analysis by preparing the data based on the selected computational algorithm in previous steps.
5. Generate original insights by organizing the evaluation of outputs in step four.
6. Present the findings by applying synthesis forms like research agenda, taxonomy, models, meta-analysis, and meta-theory to present the findings and synthesize the insights into useful building blocks for further research.

## 3. Previous Literature Works

In recent years, there has been a surge of interest in blockchain technology, leading to various research efforts in new domains, as observed by exploring databases such as Scopus and IEEE Explorer. Numerous systematic literature review (SLR) papers have been published on blockchain on security and privacy challenges. Mohanta et al. (2019) conducted an SLR on 150 articles to analyze the applications of blockchain technology, in which 20 articles discussed the challenges related to security and privacy in blockchain implementation. A study by Zhang et al. (2019) analyzed 14 articles to delve into blockchain security and privacy techniques and provides detailed descriptions of consensus algorithms, hash chained storage, mixing protocols, anonymous signatures, non-interactive ZKP, and secure MPC to achieve security and privacy in blockchain-based systems. Waheed et al. (2020) presented a technique that leverages ML to develop automated privacy and security mechanisms. Based on an analysis of 43 articles from journals and conferences, it evaluates security and privacy challenges and vulnerabilities in previous works. Ismagilova et al. (2022) reviewed 99 articles focusing on the privacy and security of mobile devices in smart cities, power systems, healthcare, and blockchain. Gugueoth et al. (2023) reviewed 31 articles on IoT, describing types of cyber attacks on IoT systems that can lead to privacy and security vulnerabilities and also explored solutions using blockchain to solve these vulnerabilities and improve IoT architecture. The SLR performed on 51 articles by Kiania et al. (2023) on the benefits of adopting blockchain in healthcare systems and its role in improving security and privacy. It also discussed the limitations of blockchain adoption in healthcare systems. A study presented by Qahtan et al. (2023) provides a comprehensive taxonomy on attribute-related security and privacy developments for blockchain-based healthcare industry 4.0. This study focused on identifying issues in healthcare industry 4.0 and addressing challenges in multi-criteria evaluation, proposing specific methods to bridge theoretical gaps. Myrzashova et al. (2023) presents an SLR with 100 articles that show how the advancement of privacy in healthcare led to the adoption of Federated Learning (FL) in healthcare systems. This study's main focus is identifying vulnerabilities in user privacy protection systems and healthcare data. Against that, the authors presented a novel conceptual framework for blockchain-enabled FL.

**Table 1. Comparison with existing literature review studies**

| Criteria | (Mohanta et al., 2019) | (Zhang et al., 2019) | (Waheed et al., 2020) | (Myrzashova et al., 2023) | (Qahtan et al., 2023) | (Gugueoth et al., 2023) | (Shahid, 2020) | Our |
|---|---|---|---|---|---|---|---|---|
| 1 | 20 | 14 | 43 | 100 | 52 | 31 | 2125 | 3904 |
| 2 | × | × | ✓ | ✓ | ✓ | ✓ | ✓ | ✓ |
| 3 | ✓ | ✓ | ✓ | ✓ | ✓ | ✓ | ✓ | ✓ |
| 4 | × | × | × | × | × | × | × | ✓ |
| 5 | × | × | × | × | × | × | ✓ | ✓ |
| 6 | × | × | × | × | × | × | × | ✓ |
| 7 | × | × | × | × | × | × | × | ✓ |
| 8 | × | × | × | × | × | × | × | ✓ |
| 9 | ✓ | ✓ | ✓ | ✓ | × | × | × | ✓ |
| 10 | × | × | × | ✓ | × | × | ✓ | ✓ |
| 11 | ✓ | ✓ | ✓ | × | × | × | ✓ | ✓ |
| 12 | × | × | × | × | × | × | ✓ | ✓ |

Note, numeric values indicate as follows: 1= number of literature included in the analysis, 2 = Quantitative Analysis, 3 = Qualitative Analysis, 4 = Data Collection Automation, 5 = Scalability, 6 = Real-Time Analysis, 7 = Algorithmic Topic Evolution Tracking, 8 = Dynamic Literature Synthesis, 9 = Trends and Pattern Identification, 10 = Unstructured Data Processing, 11 = Data-Driven Insights, 12 = Latent Topic Insight.

The literature mentioned above utilizes systematic and comprehensive literature reviews. However, it has been criticized for its inability to effectively synthesize content when dealing with a large amount of academic literature. Manual techniques and citation analysis become impractical in such cases. Shahid (2020) introduced a topic modeling-based literature review process to address this issue, using a corpus of 2,125 entries. The study employed LDA to identify topics within the corpus. The research covered practical and academic domains, with computer science being the dominant research area, followed by economics and business. This underscores the necessity for research on blockchain development across various disciplines and suggests potential future research directions. Building on this, we have adopted an unsupervised topic modeling approach to uncover trends specifically related to the security and privacy of blockchain systems. Table 1 presents the comparison between our work and previous literature work.

# 3. Methodology

This study aims to identify emerging fields and promising paths for future research related to blockchain
applications for individual users and organizations. Our methodology follows a step-by-step approach.

## 3.1. Data Collection

We started our data collections with searching articles on different databases such as IEEE, Web of Science, and Scopus, using keywords with search logic as "Blockchain" AND "Security" AND "Privacy" AND "Decentrali*" OR "Distributed". The search outcomes overlap with different databases, and most of the relevant articles are indexed in Scopus. It is also found that the collection of Scopus is comprehensive, multidisciplinary, and trusted based on abstract and citation parameters. Therefore, the Scopus database was used for conducting the CLR method through which we collected 4,975 relevant articles.

After careful filtering by subject area, article type, and language, we finally selected 4,016 articles, which we exported as a CSV file. These articles were limited to journal and conference publications in English with subject tags in Computer Science Engineering, Mathematics, Decision Sciences, Energy, Medicine, Business, Management and Accounting, Health Professions, Social Sciences, Environmental Science, Economics, Econometrics, and Finance. We meticulously removed duplicate entries and extracted preliminary information from the CSV using the versatile panda's library in Python, resulting in a refined dataset of 3,904 articles ready for further analysis. We selected the abstracts to understand the thematic pattern well and accurately identify the topics of the published papers (Principe et al., 2022).

## 3.2. Data preprocessing

In this work, data preprocessing consists of four main steps: 1) Text cleaning that includes removing unnecessary text and converting all text into lowercase which provides consistency and removes repetition of words. 2) Text terms such as "the", "that", "a", "in", etc., are identified as stop words without any significance are removed. This removal allows more emphasis to be granted to key terms with higher significance. 3) Tokenization involves splitting text into minimal considerable elements. It also applies bi-grams and trigrams on groups of frequent word pairs or triplets to treat them as single units, which improve the semantic analysis of specific sentences in the LDA model. 4) Lemmatization derives a word to its root form while preserving the context. For example, "organizational" will become "organization". In our work, we used the Spacy library for the above steps.

## 3.3. Data Processing

To better understand the research landscape within each topic, we performed an impact analysis using the litstudy (https://nlesc.github.io/litstudy/) Python library tool to extract quantitative data on the information content within our dataset. This analysis illuminated the relative influence of individual papers, authors, and journals. We calculated the impact of the paper, author, and journal based on citation count, H-index, and the impact factor of the journal (Glanzel & Moed, 2002). ¨ The litstudy library analyses textual data and calculates various impact metrics such as source of publication, year of publication, per article.

After conducting an impact analysis, we examined the content of our articles using LDA Python library Gensim (https://pypi.org/project/gensim/). LDA is capable of uncovering hidden thematic structures within the corpus. To uncover these latent themes, LDA assigns a set number of predefined topics (k) to each document iteratively, selecting the most fitting one based on the document's content. This probabilistic approach makes LDA computationally efficient, allowing us to analyze thousands of research articles effectively. We refined our corpus's textual content by eliminating punctuation, stop words (words with minimal thematic meaning), and segmenting the text into individual words (tokens). These tokens form the Document Term Matrix (DTM), a matrix that captures the frequency of each unique word within each document. We utilized Gensim's bigram and trigram functionality during tokenization to address potential ambiguities. For instance, "access" and "access control" carry distinct meanings, and this technique helps preserve these nuances within the DTM. Since LDA is unsupervised, determining the optimal number of topics (k) is crucial. We addressed this by calculating a coherence score metric. This score reflects the degree of semantic similarity between the most prominent words within each topic. Higher coherence scores indicate more interpretable and meaningful topics. Based on these scores, we selected the number of topics that produced the most semantically relevant themes. We visualized the key terms associated with each identified topic using word clouds to gain further insights. We then analyzed the most representative articles associated with each topic. This visualization provided valuable information about how blockchain technology integrates with various domains and pinpointed the most recent trending research areas in blockchain adoption.

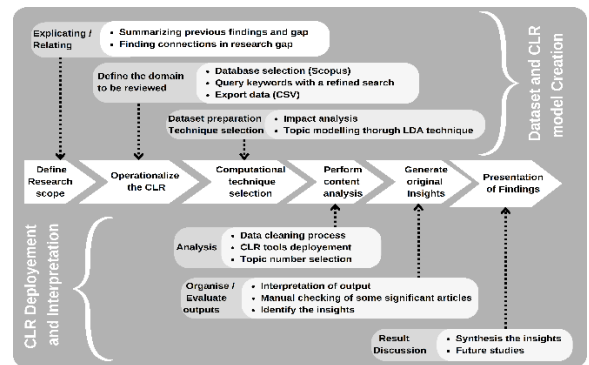

**Figure 1. Research framework**

In our research, we used the CLR method on the relevant literature, as depicted in Figure 1. The interpretation of LDA output and in-depth analysis within each topic follows an iterative cycle for optimal results. This process involves refining the analysis by adjusting parameters or diving deeper into specific topic clusters to ensure accurate and impactful conclusions.

## 4. Results

### 4.1. Impact analysis

We conducted an impact analysis on various research areas and topics based on factors such as the citation of articles, authors, and journals generated by the research community. After analyzing the data, we discovered that most of the recently published articles had little or no citations, mainly because the keywords we selected for this study were not very popular before 2016, which resulted in fewer papers to analyze. Therefore, we decided to begin our impact analysis from 2016. We have presented the number of papers published each year and how many times they were cited in Figure 2.

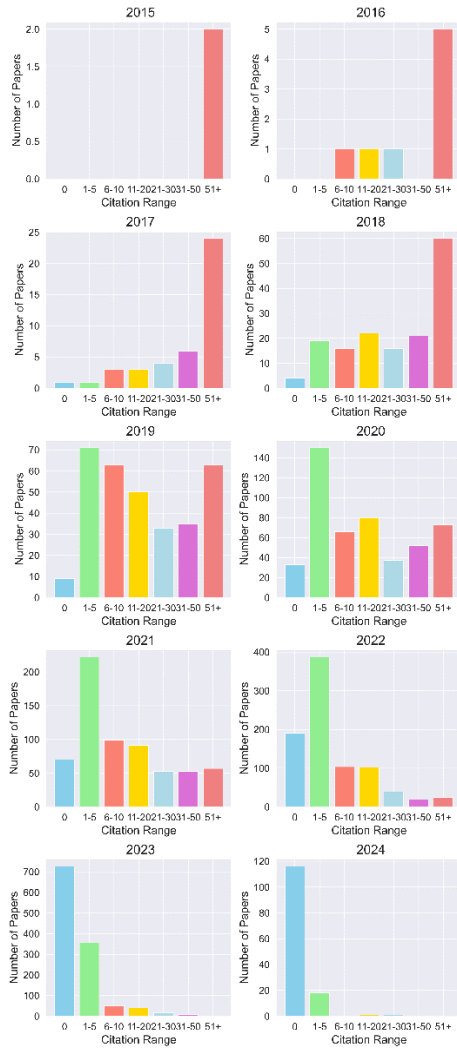

**Figure 2. Number of publications vs Citation/year**

In Figure 3, we individually present the number of publications per year to clarify the analysis of publications and citations. This visualization demonstrates the recent impacts of blockchain for information security and privacy research growth in academia. The growing interest started in 2017 and continued till 2020, when more than 1200 articles were published that predict the future research trends related to topics highlighted in different domains. In addition, we have picked out the twenty best journals and conferences based on their high number of published papers, as shown in Figure 4.

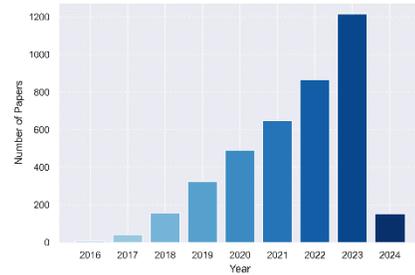

**Figure 3. Publication Per Year**

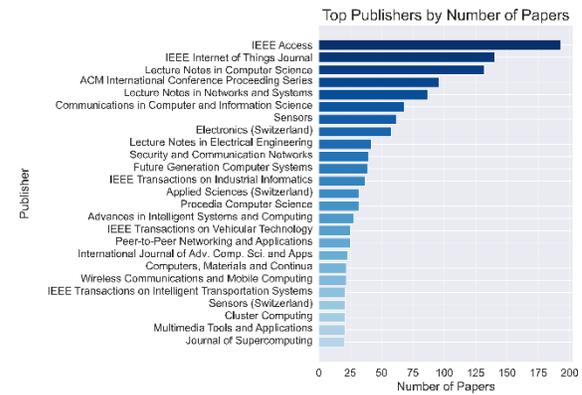

**Figure 4. Top Publishers**

In Figure 4, we found included journals from Information Systems, such as the IEEE Internet of Things Journal, IEEE Transactions on Industrial Informatics, and IEEE Transactions on Intelligent Transportation Systems. Including these IS journals Indicates That the IS encourages the advancement of security and privacy using blockchain. Also, these journals are essential in our research as they are related to developing secure and advanced infrastructure using IoT technology with the help of blockchain (Lu et al., 2019). These journals and conferences influence the impact of the selected topics

## 4.1. Content analysis

In our content analysis, we determined the number of topics and verified their accuracy by analyzing the coherence score. Figure 5 displays the coherence score of the topics from our topic modeling process. After reviewing the results in Figure 5, we chose 30 topics with a coherence score higher than 0.4 as these topics indicate high semantic relation in the corpus. Previous study also indicates that similar coherence scores produce better topic modeling outcomes (Stevens et al., 2012). These topics are visualized in a word cloud in Figure 6.

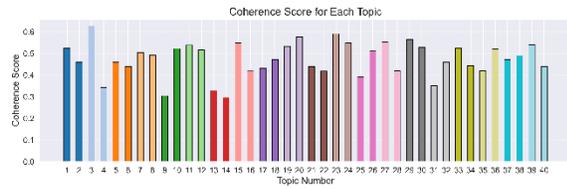

**Figure 5. Coherence Score of topics**

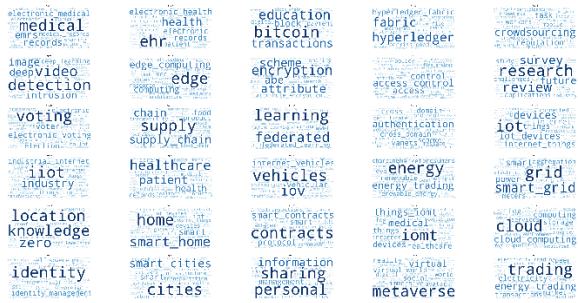

**Figure 6. Word cloud of topics**

We analyzed the word cloud and merged related topics to develop 10 new topics based on research areas for security and privacy in blockchain technology. These topics span different domains and are summarized as 1) Privacy of personal information in Healthcare, 2) Identity Management in E-education and E-voting systems, 3) Security Measures using IDS, 4) Data privacy on Cloud and Edge computing, 5) Access and authentication management in supply chain using smart contracts, 6) Advanced encryption protocols using ZKP, 7) Privacy protection in Distributed Systems using FL, 8) Security and privacy in IoT-driven systems and smart cities, 9) Data privacy in Energy trading over the smart grid, 10) Privacy of real identity in Metaverse. We have investigated to understand the evolution of topics from 2016 to 2024. In Figure 7, the topic trend indicates that the focus on blockchain-based security and privacy has increased in recent years, specifically for the supply chain, healthcare and IoT-driven systems because of the rising concern towards information security and privacy. The pattern shows the high interest of academia and industry in performing research on secure data and privacy protection in the blockchain for digital transformation in growing fields like computing, IoT, cloud, and federated learning. The surge in publications after 2020 addresses the critical role of emerging technological threats. This topic trend analysis indicates that ongoing innovation and adaptive security measures will be essential to future research directions.

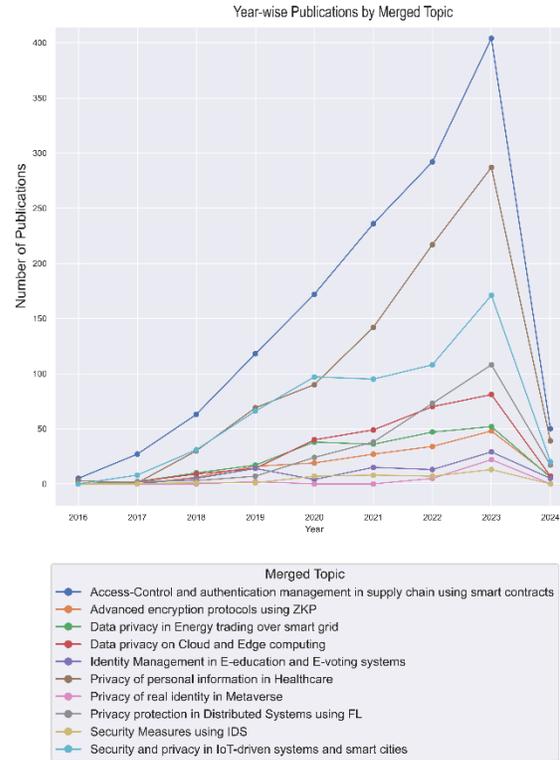

**Figure 7. The evolving topics trend over the year**

## 5. Discussion

The CLR finding provides impact and content analysis for blockchain-based information security and privacy, which helps to use resources in right direction in multidisciplinary research for Industry and academia. To clarify this, Figure 8 represents the identified topics with related keywords. We selected most impactful articles on each topic through detailed content analysis and offered an in-depth discussion with relevant sources.

T1) Privacy of personal information in healthcare: Researchers are focusing on enhancing privacy, security, and trust mechanisms through smart contracts and ZKP for self-sovereign identities. However, challenges like high implementation costs, legal issues, and complexity remain barriers to adoption (Liu et al., 2020).

T2) Identity Management in E-education and E-voting systems: Utilization of advanced cryptographic techniques in a smart contract ensures secure authentication and data management, while decentralized credentials protocols enhance privacy (Dewangan et al., 2023). However, scalability issues, economic cost and regularity hinder the integration of these advanced technologies.

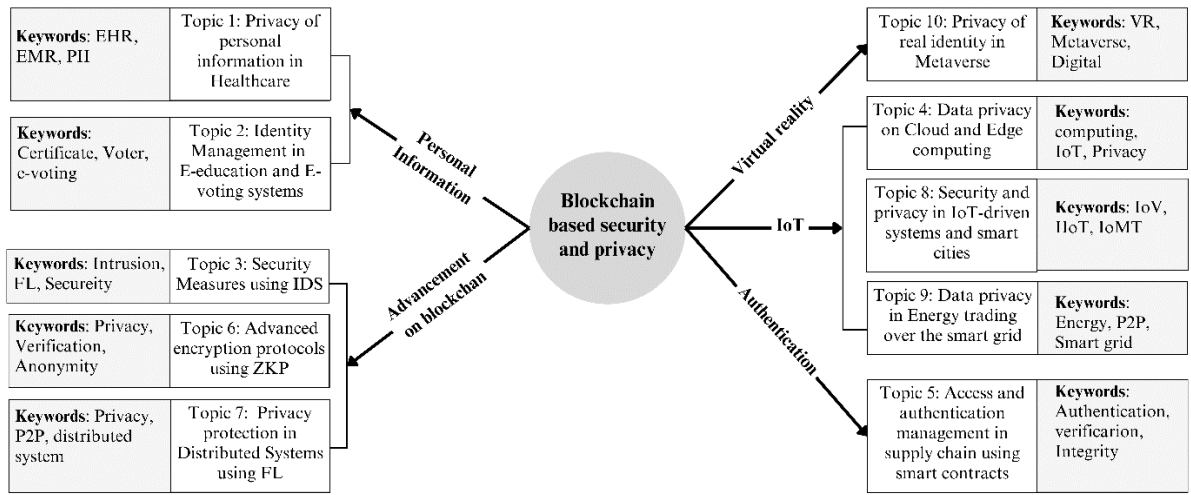

**Figure 8. Topic keyword subdivision based on security and privacy in blockchain**

T3) Security Measures using IDS: AI models can effectively detect the anomalies and security threats in blockchain-based Intrusion Detection Systems (IDS). Combining AI with ZKP can improve secure anomaly detection without compromising efficiency and data privacy. However, scalability and reliance on data quality are concerns (Saveetha & Maragatham, 2022).

T4) Data privacy on Cloud and Edge computing: Distributed authentication systems using optimized consensus and elliptic curve cryptography protect user privacy on cloud and edge platforms (Guo et al., 2019) ZKP can further enhance these systems by enabling secure user authentication without revealing sensitive information. However, Latency could be a concern due to reliance on edge nodes for authentication.

T5) Access and authentication management in the supply chain using smart contracts: Blockchain-based traceability systems use access control models to manage permissions access and data integrity in supply chains. The integration of ZKP can offer a privacy-preserving solution for verifying access rights without exposing sensitive data. However, Scalability, regulatory requirements, and the need for electronic tags remain challenging. (Dwivedi et al., 2020).

T6) Advanced encryption protocols using ZKP: these techniques ensure secure transaction verification in blockchain data without exposing private details, strengthening privacy and security (Wang et al., 2023). However, challenges like efficiency issues, the need for trusted setups, and scalability concerns still need improvement.

T7) Privacy protection in Distributed Systems using FL: Blockchain-based FL improves security and fault tolerance, handling non-independent and identically distributed data with re-encryption algorithms. However, there is a lack of cross-referencing and quantitative measures, and there is a need for broader perspectives in the future (Li et al., 2022).

T8) Security and privacy in IoT-driven systems and smart cities: Decentralized systems for IoT applications incorporate advanced cryptography and AI models that improve security and reliability. These IoT applications include the Internet of Medical Things, the Internet of Vehicles, Industrial IoT, Social IoT, and smart cities. Significant challenges in these applications include high energy consumption, maintaining trust, data integrity and scalability. (Arshad et al., 2023).

T9) Data privacy in Energy trading over the smart grid: adoption of blockchain in smart grids enabling privacy protection, identity authentication, data aggregation, and electricity pricing. However, scalability, computational complexity, privacy vulnerabilities, integration challenges, and energy consumption remain limitations of these systems (Cao et al., 2023).

T10) Privacy of real identity in Metaverse: Blockchain integration addresses security challenges like injection attacks and user profiling. Additionally, ZKP ensures strong authentication without compromising user privacy (Huang et al., 2023). However, integration complexity, regulatory concerns, and scalability remain challenges for the wide adoption of this technology.

Based on our content and impact analysis findings, we have developed future research questions. These questions can translate into blockchain research opportunities to protect privacy and security.

T1) Privacy of personal information in healthcare: How can blockchain enhance secure personal data and identity verification in an interoperable healthcare network?, How does federated learning help utilize patient data while maintaining privacy?

T2) Identity Management in E-education and E-voting systems: How can blockchain be enhanced for better security and privacy in e-voting with quantum computing?, How will advanced protocols like ZKP, MPC and FL improve privacy in e-voting ?

T3) Security Measures using ID: How can blockchain-based IDS be optimized for real-time cyber threat detection using FL?, How can IDS system scalability be ensured while maintaining high security?

T4) Data privacy on Cloud and Edge computing: How can ZKP and FL improve data privacy in cloud and edge computing?, What methods can mitigate attack risks in edge computing environments?

T5) Access and authentication management in the supply chain using smart contracts: What factors affect blockchain adoption's scalability and energy consumption, which it is solved by FL? Is there a need for ZKP to protect privacy, new access control, and authentication mechanisms in blockchain?

T6) Advanced encryption protocols using ZKP: How can ZKP systems be more scalable for widespread adoption?, What techniques can emerge to reduce the high energy consumption of ZKPs in blockchain?

T7) Privacy protection in Distributed Systems using FL: What techniques exist to improve the precision and selectivity of federated learning models in decentralized systems?, How can the trust-establishing characteristics of blockchain be used to enhance federated learning frameworks?

T8) Security and privacy in IoT-driven systems and smart cities: How can we simplify data processing process ny integrating FL in IoT infrastructure?

T9) Data privacy in Energy trading over the smart grid: How can a smart grid balance the trade-off between privacy preservation and communication complexity?, Is FL suitable for optimizing energy trading in the smart grid using blockchain while preserving privacy?

T10) Privacy of real identity in Metaverse: What prevention mechanisms should be developed to stop hackers from exploiting vulnerabilities in the metaverse? How can ZKP and MPC enhance authentication processes and protect users' real identities in virtual environments?

Analysis of the topics and proposed research questions show that blockchain has a high potential to address identity management and data security. However, topics like healthcare, online education, supply chain and metaverse are highly concerned about protecting privacy in the blockchain. Although previous literature has addressed this issue by integrating advanced security protocols, it also highlights challenges in practical implementations related to scalability, interoperability, global standardization and insecurity from advanced and powerful computing technologies. However, the integration of advanced security and privacy methods like ZKP, MPC, and FL are rising with claims that they can address these challenges that hinder their wide adoption.

## 6. Contributions

From the theoretical perspective, our study is one of the first studies conducted in IS that applied the CLR technique. Based on our literature review findings, we discussed issues related to information security and privacy in the IS domain. These findings show that blockchain has been widely implemented in industry and public sectors such as healthcare, education, and supply chains. Further, our findings suggest that blockchain-based systems can enhance information security and privacy management using methods such as FL, ZKP, and MPC. Future IS research can investigate the implications of these techniques in terms of enhancing security and privacy in various industry and public sectors.

Our findings provide practical implications of blockchain-based secure and privacy-oriented frameworks that process and store users' personal information. These frameworks can adopt ZKP protocol-based operations on encrypted data in an interoperable environment that keeps original data secure during analysis. Additionally, integrating FL and IoT provides scalability and supports interoperability by processing data at local devices. These systems are adaptable for both industry and public services, where handling users' personal information and data is crucial. As these systems can perform intelligent and decentralized data processing with less energy consumption, they can also provide a sustainable solution.

## 7. Limitations of the work

However, our study is not without any limitations. Using the LDA requires a predetermined number of topics (K) to prevent overfitting or underfitting. In LDA, subjective evaluation of the analysis impacts the resulting topics. Due to feeding content in different orders, LDA may produce varied outcomes, leading to temporal biases and confusion. Furthermore, relying on only one database may sometimes result in limited article search results.

## 8. Conclusion

Use We conduct CLR on a corpus of 3904 documents and explored privacy for information management. Through impact and content analysis, we identified the journals and conferences that support our chosen area and the number of past publications. LDA topic modelling was used to identify 10 trending topics within the corpus. We manually reviewed literature related to these topics to ensure our findings were accurate. Our results and discussions showed that blockchain technology significantly impacts privacy-preserving techniques and secure information management across various domains. We also present the evolving topics trend over the year to show the growing interest in blockchain-based security and privacy research. We focused on emerging topics and suggested possible future Information Security and Privacy Protection research opportunities.

## 9. Acknowledgement

This work was supported by the Research Council of Finland with CHIST-ERA, grant agreement no - 359790, Di4SPDS-Distributed Intelligence for Enhancing Security and Privacy of Decentralized and Distributed Systems.